\pdfoutput=1
\documentclass[final,5p,times,twocolumn,a4paper]{elsarticle}
\usepackage{graphics}
\usepackage{amssymb}
\journal{Physica C}

\begin{document}

\begin{frontmatter}

\title{Nuclear Magnetic Relaxation Rate in Iron-Pnictide Superconductors}

\author{T.~Kariyado}
\ead{kariyado@hosi.phys.s.u-tokyo.ac.jp}
\author{M.~Ogata}
\address{Department of Physics, University of Tokyo, Hongo,
 Bunkyo-ku, Tokyo 113-0033, Japan}
\address{JST. TRIP, Sanbancho, Chiyoda, Tokyo 102-0075, Japan}

\begin{abstract}
Nuclear magnetic relaxation rate $1/T_1$ in iron-pnictide
 superconductors is calculated using the gap function obtained in a
 microscopic calculation. Based on the obtained results, we discuss
 the issues such as the rapid decrease of $1/T_1$ just below the
 transition temperature and the difference between nodeless and nodal
 s-wave gap functions. We also investigate the effect of Coulomb
 interaction on $1/T_1$ in the random phase approximation
 and show its importance in interpreting the experimental results.
\end{abstract}


\end{frontmatter}

Iron-pnictide superconductors have been one of the central issues in
condensed matter physics since its discovery. Many experimental and
theoretical works have been carried out to elucidate the symmetry and
the
structure of the superconducting gap function since the information on
the
gap structure gives a vital clue to determine the origin of the
superconductivity. Recently, it is recognized that the 
iron-pnictide superconductors show varieties of the structure of the gap
function\cite{Hashimoto:2009}, i.e., nodeless behavior in some systems
and nodal behavior in other systems.

NMR measurement of $1/T_1T$ is useful in determining such detailed
structure of the gap function and many experimental results have been
published\cite{Yashima:2009,Matano:2008,Kobayashi:2009,Fukazawa:2009a,Nakai:2009}.
Thus, we calculate
$1/T_1T$ based on some models for iron pnictides and compare the obtained
results with experiments. Since many theoretical works
exist\cite{Nagai:2008,Parker:2008a,Bang:2008,Parish:2008,Seo:2009}, we
concentrate on the unresolved issues such as the rapid decrease of
$1/T_1$ just below the transition temperature
and the discussion on the difference between nodeless and nodal s-wave
gap functions.

For this purpose, we use the five-orbital Hubbard model proposed in ref
\cite{Kuroki:2009}. Since we want to discuss the difference between the
nodeless s-wave and nodal s-wave, we use following two models. One is
a simplified model downfolded from the LDA calculation on
the LaFeAsO system. Here ``simplify'' means that we neglect the three
dimensionality and take hopping integrals only up to the fifth nearest
neighbors. Other is a three dimensional model of the LaFePO
system. All results shown below are obtained with band filling $n=6.1$.

Assuming that the coupling between the conduction electrons and nuclear
spins is diagonal in the orbital basis and using the constant form
factor $A(\vec{q})=1$, $1/T_1T$ can be written as
\begin{equation}
 \frac{1}{T_1T}=\lim_{\omega\rightarrow 0}
  \frac{1}{N}\sum_{\vec{q}}\sum_{ab}
  \frac{\mathrm{Im}\chi_{aabb}(\vec{q},\omega)}{\omega}.
\end{equation}
Note that $a$ and $b$ denote the orbital indices. Without Coulomb
interaction, we have
\setlength{\arraycolsep}{0mm}
\begin{eqnarray}
 \chi_{abcd}(q)=&&\nonumber\\
 -\frac{T}{N}\sum_{q'}&&
  \left\{G_{ac}(q+q')G_{db}(q')+\bar{F}_{bc}(q+q')F_{da}(q')\right\}
  \label{chi0}
\end{eqnarray}
where $q=(i\omega_n,\vec{q})$ and $G_{ab}(q)$ ($F_{ab}(q)$) is the
normal (anomalous) Green's function in the orbital representation. We
also take acount of the Coulomb interaction within the
random phase approximation (RPA), where the enhancement
factor $(\hat{1}-\hat{\chi}(q)\hat{V}^s)^{-1}$
is included. Note that $\hat{\chi}(q)$ is the matrix form of
eq.~(\ref{chi0}) and $\hat{V}^s$ is the spin vertex defined as
$V^s_{abcd}=U\ (a=b=c=d)$, $U'\ (a=c,b=d,a\neq b)$, $J_H\ (a=b,c=d,a\neq c)$,
$J\ (a=d,b=c,a\neq b)$, 0 (others). In what follows, we use $U=1.0$ eV
, $J_H=0.2$ eV, $U=U'+J_H+J$ and $J_H=J$. 

As is noted above, we describe the coupling between electrons and
nuclear
spins and the Green's function in the orbital basis. Thus, we
describe the gap function also in the orbital basis. In this orbital
representation, the gap function becomes matrix form as
$\hat{\Delta}(\vec{k})$. Limiting the discussion on the s-wave channel,
physically important entries of $\hat{\Delta}(\vec{k})$ can be written as
\begin{eqnarray}
 \Delta_{22}(\vec{k})&=&a_2+2b_2(\cos k_x+\cos k_y)
  +4c_2\cos k_x\cos k_y\nonumber\\
  &&\ \ \ \ \ \ \ \ \ \ \ \ \ \ \ \ \ \ \ \ \ \ \ 
   \ \ \ \ \ \ \ \ \ \ \ \ \ \ \ -4\delta \sin k_x\sin k_y,\label{d22}\\
 \Delta_{33}(\vec{k})&=&a_2+2b_2(\cos k_x+\cos k_y)
  +4c_2\cos k_x\cos k_y\nonumber\\
  &&\ \ \ \ \ \ \ \ \ \ \ \ \ \ \ \ \ \ \ \ \ \ \ 
   \ \ \ \ \ \ \ \ \ \ \ \ \ \ \ +4\delta \sin k_x\sin k_y,\label{d33}\\
 \Delta_{44}(\vec{k})&=&a_4+2b_4(\cos k_x+\cos k_y)
  +4c_4\cos k_x\cos k_y, \label{d44}\\
 \Delta_{23}(\vec{k})&=&2d_{23}(\cos k_x-\cos k_y),\label{d23}\\
 \Delta_{24}(\vec{k})&=&2ip_{24}(\sin k_x-\sin k_y).\label{d24}
\end{eqnarray}
Note that subscript 2, 3 and 4 represent $zx$, $yz$ and $x^2-y^2$
orbital
respectively. To determine the parameters in eqs.~(\ref{d22}-\ref{d24}),
we solve the linearized Eliashberg equation with the effective
interaction calculated within RPA (we follow the formalisms in
ref~\cite{Kuroki:2009}). Then, we obtain the gap function with interband
sign reversal witch can be parameterized as ($a_2$, $b_2$, $c_2$,
$\delta$,
$a_4$, $b_4$, $c_4$, $d_{23}$, $p_{24}$) = (0.0075, -0.038, -0.0155,
0.0075, 0.21, 0.11, -0.047, -0.0054, 0.045) for LaFeAsO and (0.060,
-0.030, -0.011, 0.0050, -0.072, 0.024, -0.0044, -0.022, -0.017) for
LaFePO. The important difference between the two results is that 
$\Delta_{44}(\vec{k})$ has weaker $\vec{k}$ dependence in LaFePO and
this weak dependence leads to the node when the gap function is written
in the band representation\cite{Kuroki:2009}. Namely, the gap function
is nodeless in LaFeAsO and nodal in LaFePO. 

For studying the gap function below $T_c$, we assume the usual
temperature dependence, i.e., we use
\begin{equation}
 \hat{\Delta}(T,\vec{k})=\alpha \hat{\Delta}(\vec{k})
  \tanh\left(1.74\sqrt{T_c/T-1}\right).\label{deltaT}
\end{equation}
In this work, $T_c$ is fixed to a large value of 0.02 eV to avoid the
numerical difficulty. Scaling factor $\alpha$ in eq.~(\ref{deltaT}) is
determined to have $2\Delta^{\mathrm{max}}/k_BT_c\sim 8$ that is typical
for iron-pnictide superconductors with $\Delta^{\mathrm{max}}$ being the
largest gap value on the Fermi surface.

\begin{figure}[t]
 \begin{center}
  \includegraphics[width=5.8cm]{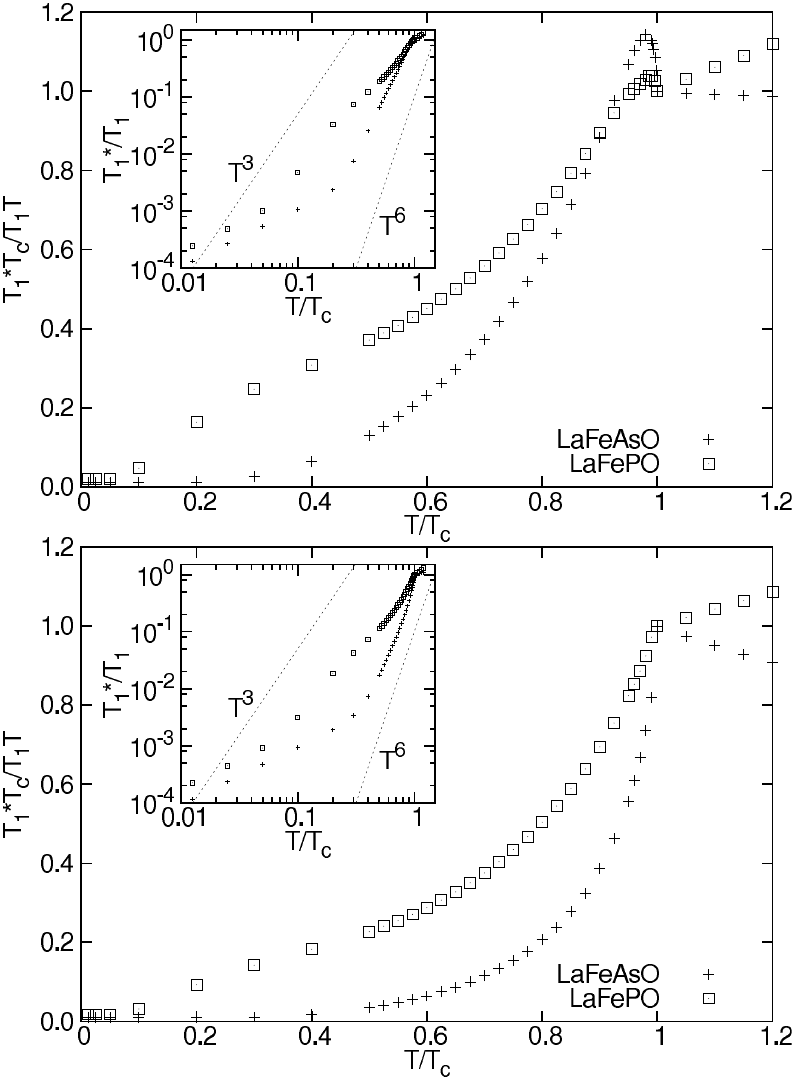}
  \caption{$T_1^*T_c/T_1T$ without Coulomb interaction where $T_1^*$
  is the $T_1$ value at $T_c$. Calculation is done
  with 192$\times$192 (100$\times$100$\times$8) meshes in the momentum
  space for LaFeAsO (LaFePO) system. Upper (lower) panel shows the results
  without (with) Coulomb interaction.}
  \label{fig1}
 \end{center}
\end{figure}
First, we show the results without Coulomb interaction in the 
upper panel of Fig.~\ref{fig1}. In this calculation, we use the damping
factor $\gamma=0.05T_c$ to discuss the clean limit ($\gamma$ is defined
as $i\omega_n\rightarrow \omega+i\gamma$). Although the used
gap functions have unconventional property (interband sign reversal),
the results show that the small coherence peaks appears in both of
LaFeAsO and LaFePO system. (The coherence peak in LaFePO is neglegibly
small.)
Inset of the upper panel of
Fig.~\ref{fig1}
shows that $1/T_1$ decreases slower in LaFePO system that has the nodal
gap function compared with LaFeAsO system as is expected. Actually,
$1/T_1$ in LaFePO decreases slower than $T^3$ just below $T_c$.

Next, we show the results with Coulomb interaction treated within RPA in
the lower panel of 
Fig.~\ref{fig1}. The most important difference between results with and
without Coulomb interaction is the disappearance of the coherence peaks.
Here, we use the same damping factor $\gamma=0.05T_c$ as in the case of
without Coulomb interaction, i.e., we assume the same
strength of the impurity scattering. This means that we can eliminate
the coherence peaks without including the impurity effects. The
reason for this disappearance
is that Coulomb interaction makes the interband contribution to $1/T_1$
more dominant than the intraband contribution and the dominance of
the interband contribution leads to the disappearance of the coherence peak
for the gap function with an interband sign
reversal\cite{Parker:2008a,Seo:2009}. Another important point is that 
Coulomb interaction gives faster decrease of $1/T_1$ just below $T_c$ as
can be seen by
comparing the inset of upper and lower panel of Fig.~\ref{fig1}.
Paying attention to the temperature just below $T_c$, the
result for LaFeAsO system shows more rapid decrease than $T^6$, which is
the experimintally claimed rapid decrease\cite{Kobayashi:2009},
and the result for LaFePO
system shows the decreasing rate close to $T^3$. Note that such details
depend on the value of $\alpha$ as is the case without Coulomb interaction.
 
In summary, we have investigated the difference of the nodeless and
nodal s-wave
gap function through calculating $1/T_1T$ using the microscopically
determined gap functions. We have also demonstrated that
Coulomb interaction can be a possible reason for the non existence of
the coherence peaks in the clean limit. Further, we have shown the
importance of Coulomb interaction in the discussion of the power of
$1/T_1$ just below $T_c$.

\section*{Acknowledgment}
We thank K.~Kuroki for useful comments and for sharing the data of
downfolded models.

\end{document}